\documentstyle[aps,preprint,eqsecnum]{revtex}
\tighten

\begin{document}
\draft

\preprint{HD-09}

\title{Response of a Kinetic Ising System to Oscillating \\
External Fields: Amplitude and Frequency \\
Dependence}

\author{S.W. Sides,$^{\ast \dagger}$
R.A. Ramos,$^{\ast \dagger}$ P.A. Rikvold,$^{\ast \dagger}$
and M.A. Novotny $^{\dagger \ddagger}$}

\address{$^{\ast}$Center for Materials Research and Technology and Department
of Physics,\\
$^{\dagger}$Supercomputer Computations Research Institute, \\
Florida State University, Tallahassee, Florida 32306-3016 \\
and \\
$^{\ddagger}$ Department of Electrical Engineering \\
FAMU/FSU College of Engineering, Tallahassee, Florida 32310 \\}

\maketitle

\begin{abstract}
The $S$=$1/2$, nearest-neighbor, kinetic Ising model has been
used to model magnetization switching in nanoscale ferromagnets.
For this model, earlier work based on the droplet theory of
the decay of metastable phases and Monte Carlo simulations has shown
the existence of a size dependent
spinodal field
which separates deterministic and stochastic
decay regimes.
We extend the above work to study the effects of an
oscillating field on the magnetization response
of the kinetic Ising model.
We compute
the power spectral density
of the time-dependent magnetization
for different values of the amplitude
and frequency of the external field, using Monte Carlo simulation data.
We also investigate the
amplitude and frequency dependence of the
probability distributions for the  hysteresis loop area and
the period-averaged magnetization.
The time-dependent response of
the system is classified by analyzing
the behavior of these quantities within the framework of
the distinct deterministic and stochastic decay modes mentioned above.
\end{abstract}


\narrowtext

\bibliographystyle{unsrt}

\section{Introduction}

Hysteresis is a nonequilibrium phenomenon characteristic of metastable systems.
The kinetic Ising model, for temperatures below its
critical temperature $T_{c}$ and at non-zero external fields $H$,
exhibits a metastable phase. If $H(t)$ varies periodically in time,
the response of the system, the magnetization $m(t)$, lags behind the
forcing field, and hysteresis occurs. Previous studies of hysteresis
have been performed, both for mean-field models
and for the Ising model \cite{rao,muktishE}.
In both cases the hysteresis loop area, $A$=$-\oint m dH$,
was found to have a
power-law dependence on the
frequency and amplitude
of $H$ for low frequencies.
Also, the mean-field models exhibit
a dynamic phase
transition in which the period-averaged magnetization,
$Q$=$(\omega / 2 \pi) \oint m dt$, changes from
$Q\neq0$ to $Q$=$0$ \cite{tome,lo}.
Magnetization reversal in small
ferromagnetic grains has been modeled with
a kinetic Ising model in static fields \cite{howard}.
Also, recent experiments on ultrathin ferromagnetic
Fe/Au(001) films \cite{hysExpt1} have considered
the frequency dependence of
hysteresis loop areas, obtaining exponents that are consistent with those
found for the two dimensional Ising model \cite{rao}.

The model used in our study is a kinetic, nearest-neighbor
Ising ferromagnet on a
square lattice with periodic boundary conditions.
The Hamiltonian is given by
${\cal H }$=$-J \sum_{ {\em \langle ij \rangle}} {\em s_{i}s_{j}} - H \sum_{i}
{\em s_{i}}$,
where $\sum_{ {\em \langle ij \rangle} }$ runs over all
nearest-neighbor pairs, and $\sum_{i}$ runs over all
$N$=$L^{2}$ lattice sites.
The dynamic used in this work
is the Glauber \cite{glauber1} single spin-flip Monte
Carlo algorithm.
The system is put in contact with a heat bath at temperature
$T$, and each
spin can flip
from ${\em s_{i}}$ to ${\em -s_{i}}$ with a
probability given by \cite{glauber2}
\begin{equation}
W(s_{i} \rightarrow -s_{i}) = \frac{ \exp(- \beta \Delta E_{i})}{1 + \exp(-
\beta \Delta E_{i})} \ ,
\end{equation}
where $\Delta E_{i}$ gives the change in the energy of the system
if the spin flip is accepted,
and $\beta = 1/k_{\rm B}T$.

In contrast to earlier studies of hysteresis in Ising models, we have
performed simulations in two distinct field regimes.
It has recently been observed that the metastable phase in Ising
models exposed to a static field $H$ decay by different mechanisms,
depending on $H$ and the system size $L$ \cite{rtms}.
For fields weaker than
a $T$ and $L$ dependent dynamic
spinodal field $H_{\rm DSP}$, the mean lifetime $\tau$
of the metastable phase
is comparable to the standard deviation in the lifetime.
This field region is termed the stochastic or single-droplet (SD)
region because decay of the metastable phase proceeds by random
nucleation of a single critical droplet of the stable phase.
For fields stronger than
$H_{\rm DSP}$, the mean lifetime of the metastable phase
is much greater than the standard deviation in the lifetime.
This field region is called the deterministic or
multi-droplet (MD) region because
decay of the metastable phase proceeds by the nucleation, growth, and
coalescence of many droplets of the stable phase.
The study presented here is, to our knowledge, the first in which the
effects of these two different decay mechanisms on hysteresis
are considered.

In our simulations, a system of size $L$ at temperature $T$
is prepared  with $m(0)$=$0$ and the up and down spins in
random arrangement.
Then, a sinusoidal field $H(t) = H_{o}\sin (\omega t)$ is applied,
and the magnetization $m(t)$ is recorded.
The field $H(t)$ is changed every attempted
spin flip, allowing for a smooth variation.
Using the $m(t)$ vs.\ $t$ data, we calculate power spectral densities
(PSD) and probability distributions for the loop area $A$ and
the period-averaged magnetization $Q$ for different amplitudes
and frequencies of $H(t)$.

\section{Results}

Our results are presented so as to contrast
the behaviors of the hysteretic
magnetization response in the SD and MD regions.
All of the results shown are for
$L = 60$ and $T$=$0.8T_{c}$.
The simulations were performed
at two field amplitudes $H_{o}$ and several frequencies $\omega$.
One value of $H_{o}$ was
chosen such that for $\omega=0$, the system was clearly in the SD
region.
Similarly, the other value of $H_{o}$ was chosen such that the system
was definitely in the MD region.
While several frequencies were used,
here we show results for a value of $\omega$ for which
the magnetization switches many
times over the course of the entire simulated
time series (although not
necessarily with the same periodicity as $H(t)$).
Figure 1 shows a portion of the time series in the MD regime,
where the system
switches over several periods
from a state that oscillates with $Q<0$, to a state
that oscillates with $Q>0$.
In Fig.\ 2, a portion of
the time series in the SD regime
shows oscillations with $Q$ near plus or minus the
zero-field spontaneous magnetization, punctuated by random
switching events that are completed within less than one period
of the field oscillation.
Here, once a critical droplet forms (in a Poisson process),
the stable phase quickly takes
over the system.

Figure 3 shows the PSDs for the entire
time series of Figs.\ 1 and 2.
The first peak in the MD data is at the
driving frequency
$f$=$(\omega/2 \pi)$ of the external field.
The higher-frequency peaks are located at integer
multiples of $f$.
For the SD data the resolution at low frequencies
is not fine enough to show even the first peak. However, other data in the
SD regime using $H(t)$ with shorter periods do show
a peak at the driving frequency.
Longer time series are needed to resolve the low-frequency
part of the spectrum.
Both the MD and SD data show large low-frequency
components, indicating
the slow dynamics of the magnetization reversal in both regimes.
In the high-frequency part of the spectrum, the PSDs
have slopes on the log-log
plot of approximately two, corresponding to exponential short-time
correlations.

Figure 4 shows the probability distributions of the hysteresis loop areas
$A$ for the data of Figs.\ 1 and 2.
The distribution for the MD regime shows a single wide
peak, since $m(t)$ lags
behind $H(t)$ by a nearly constant phase factor.
The distribution in the SD regime
has a narrow peak near zero as $m(t)$ oscillates near
the positive or negative
spontaneous magnetization for most of the length of the run.
The long tail at higher $A$ values corresponds to the occasional
rapid switching events.

Figure 5 shows the distributions of the period-averaged
magnetization $Q$
for the data in Figs.\ 1 and 2.
The distributions for both the SD and
the MD regions show a double-peaked structure. The two sharp peaks in
the SD data are due to $m(t)$ oscillating
near the spontaneous magnetization values
during most of the run.
The distribution for the MD regime shows two peaks as well, but
each peak is much wider than in the SD case.
The positions of the peaks in both distributions have been found to
be frequency dependent as well as amplitude dependent. In both cases
the asymmetry of the distribution is an effect of the finite length
of the time series.

\section{Conclusion}
Our results show distinct differences between the
multi-droplet region and the single-droplet region. This study
shows that the nature of the response of an Ising system depends not
only on the competition between the two time scales: the oscillation
period of the external field, and the lifetime of the metastable state.
The response also depends on the mode by which the system switches
magnetization states, which could depend not only on the field amplitude
and frequency, but also on the temperature and system size.
In future studies, we plan to use
longer time series to obtain better low-frequency
resolution in the PSDs and better statistics for the
probability distributions of $A$ and $Q$.

\acknowledgments

Supported in part by
FSU-MARTECH,
by
FSU-SCRI under DOE
Contract No.\ DE-FC05-85ER25000,
and by NSF
Grants No.\ DMR-9315969 and DMR-9520325.


\begin{figure}
\caption{Magnetization $m(t)$ and external field $H(t)$
vs.\ time $t$ in the MD region with
$(2 \pi/ \omega)=276$ Monte Carlo steps per spin (MCSS),
$H_{o}=0.3J$, and the lifetime in static field
$\tau_{\omega=0} \approx 55$ MCSS.
The darker line shows $m(t)$ and the lighter line
denotes $H(t)$.
The total length of the time
series is approximately $2\times10^{\bf 5}$ MCSS. The ratio of the period
of $H$ to $\tau_{\omega=0}$ is approximately $5$.}
\end{figure}

\begin{figure}
\caption{Magnetization $m(t)$ and external field $H(t)$
vs.\ time $t$ in the
SD regime with $(2 \pi/ \omega)=5000$ MCSS,
$H_{o}=0.1J$, and
$\tau_{\omega=0} \approx 1000$ MCSS.
The darker line shows $m(t)$ and the lighter line
denotes $H(t)$.
The total length of the time series is approximately
$4\times10^{\bf 5}$ MCSS.
The ratio of the period of $H$ to
$\tau_{\omega=0}$ is approximately $5$.}
\end{figure}

\begin{figure}
\caption{Power spectral densities (PSDs)
for the time series of Figs.\ 1 and 2. The magnetization is sampled
every 0.1 MCSS, so the Nyquist frequency is $5\ {\rm MCSS}^{-1}$.
The lowest frequency that can be resolved for both PSDs is
$0.2 \times 10^{-4}\ {\rm MCSS}^{-1}$.}
\end{figure}

\begin{figure}
\caption{Probability distributions of the hysteresis
loop areas,
$A$=$-\oint m dH$.
$A$ is calculated
for the MD ($\bullet$) and
SD ($\Box$) regions from the data of Figs.\ 1 and 2.
We have used bins of size $0.1$ for
the area values for both data sets. Note that the loop areas have been
normalized by $4H_{o}$, the maximum possible loop area.
The number of events (periods)
for the SD data and MD data is different; the MD distribution uses
$759$ events and the SD distribution uses $83$ events.}
\end{figure}

\begin{figure}
\caption{Probability distributions of the period-averaged
magnetization, $Q$=$ (\omega /2 \pi) \oint m dt$. $Q$ is
calculated for the MD and
SD regions from the data in Figs.\ 1 and 2.
The size of the bins used and
the quality of the statistics are the same as in Fig.\ 4.}
\end{figure}

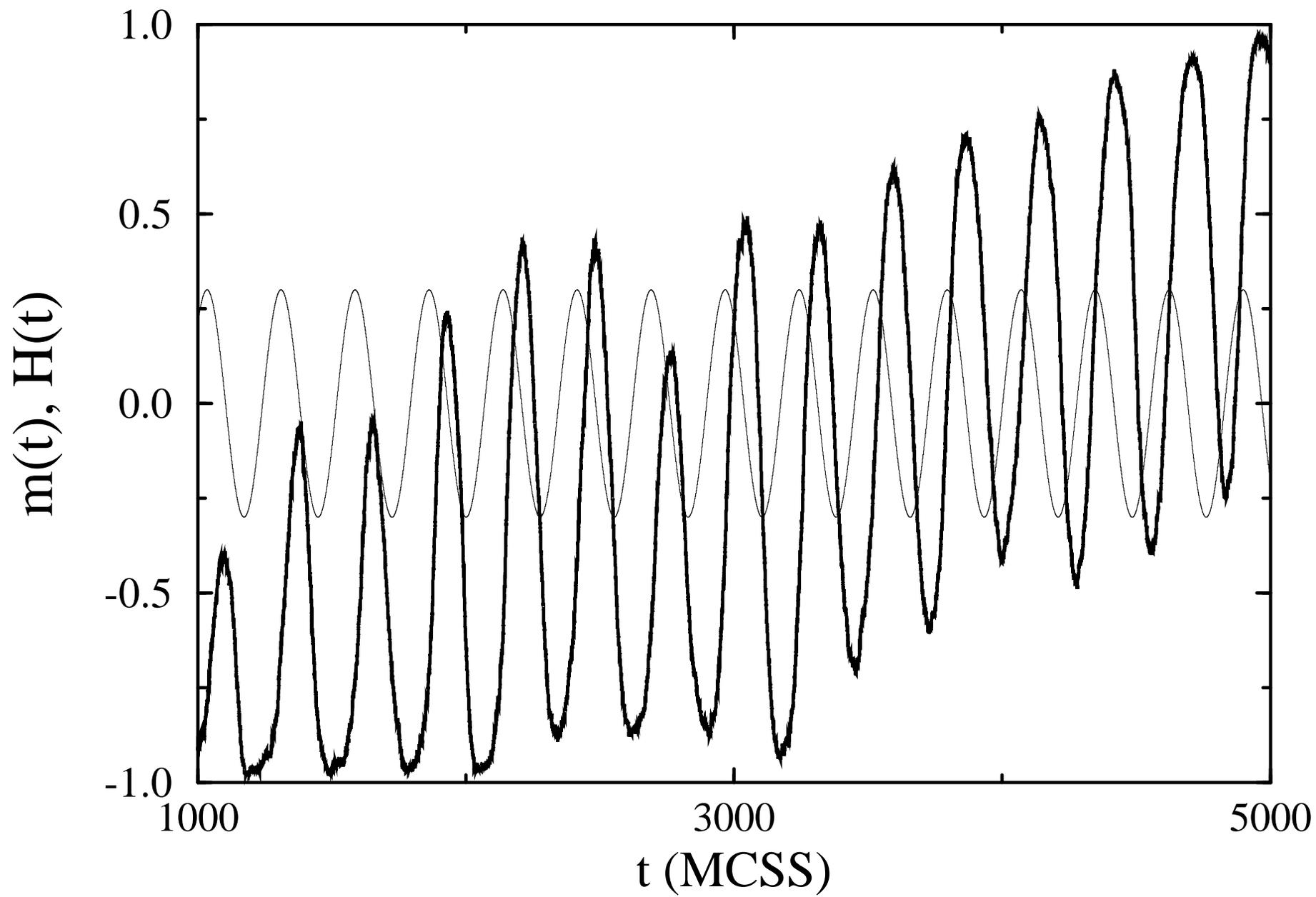
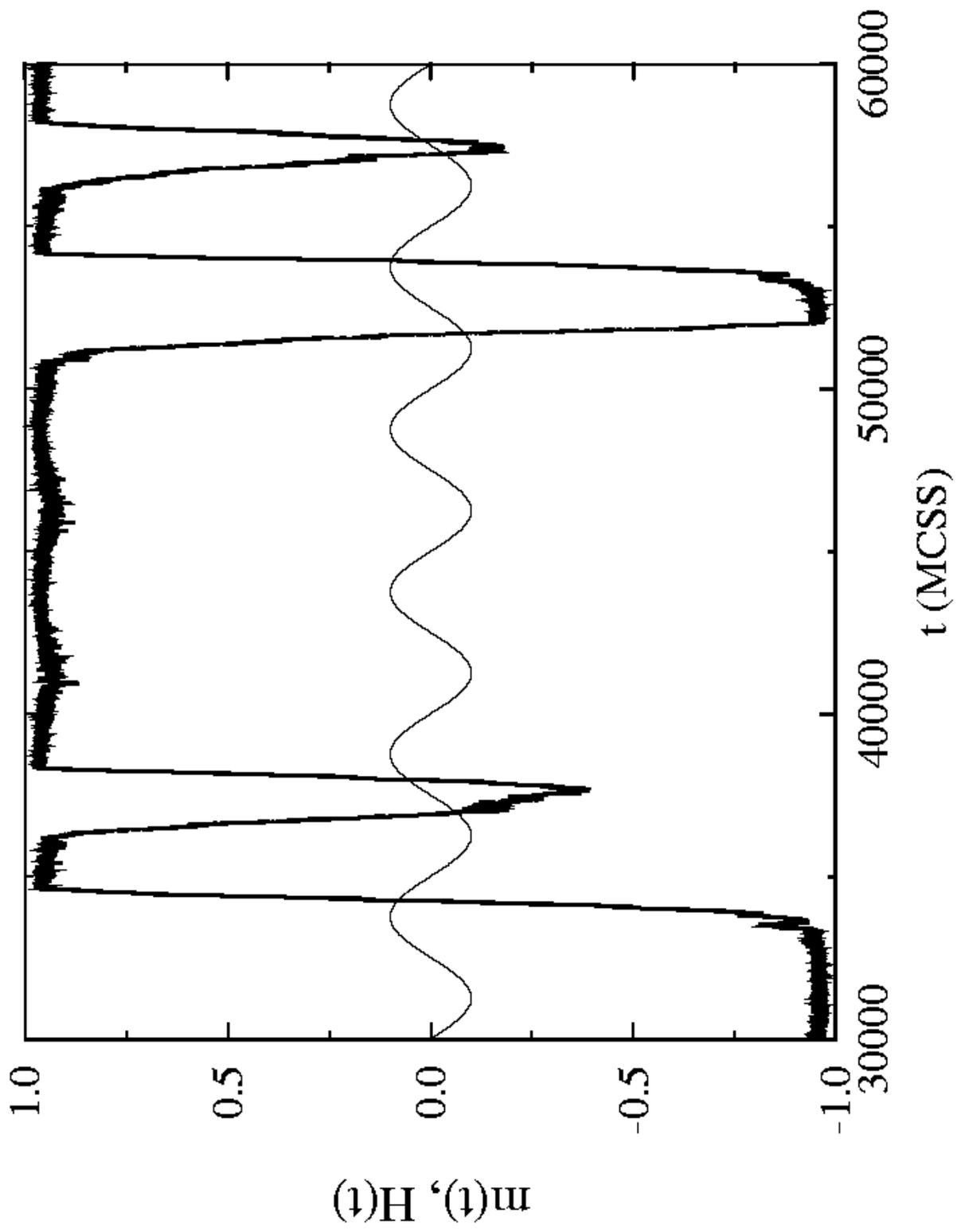
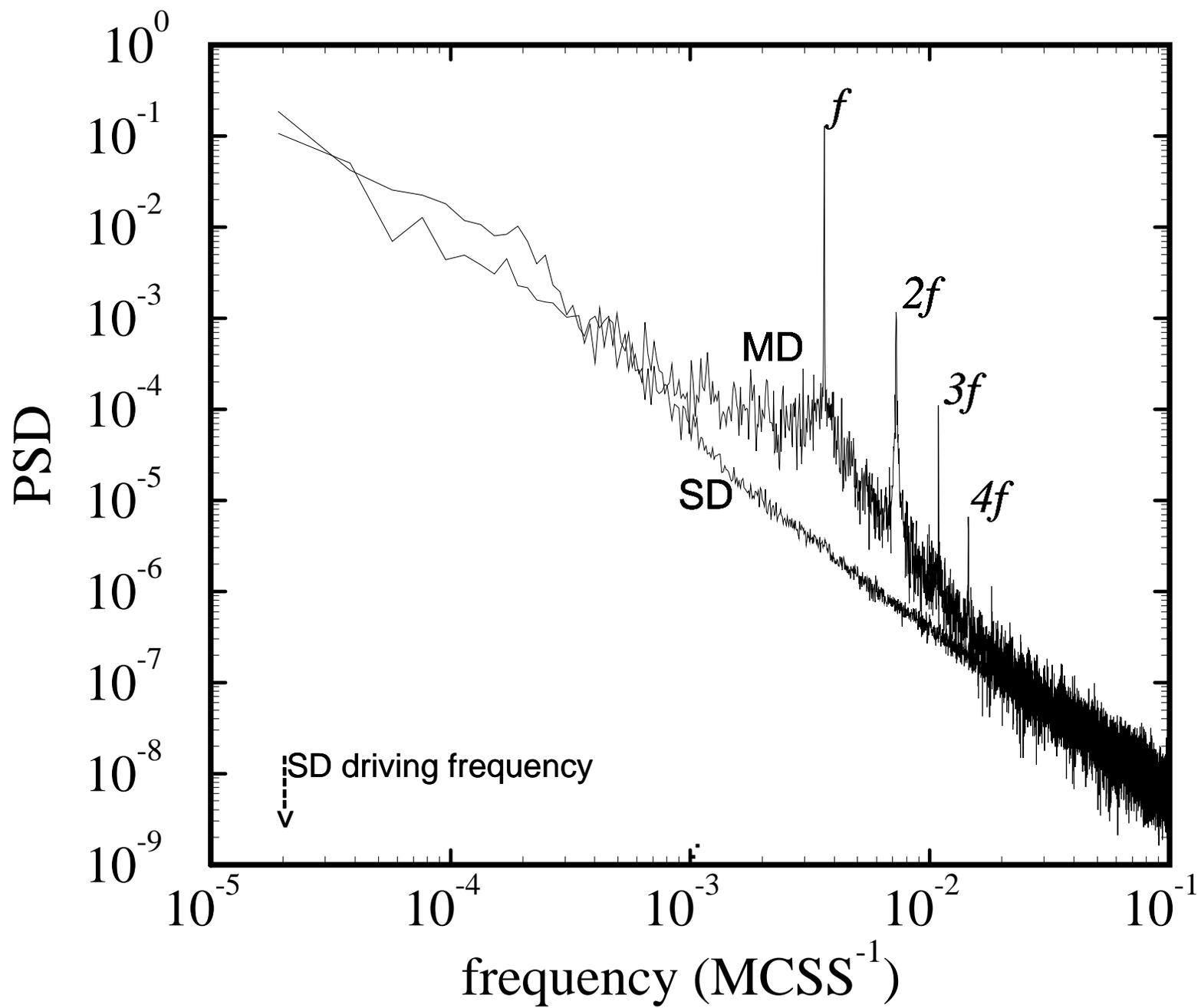
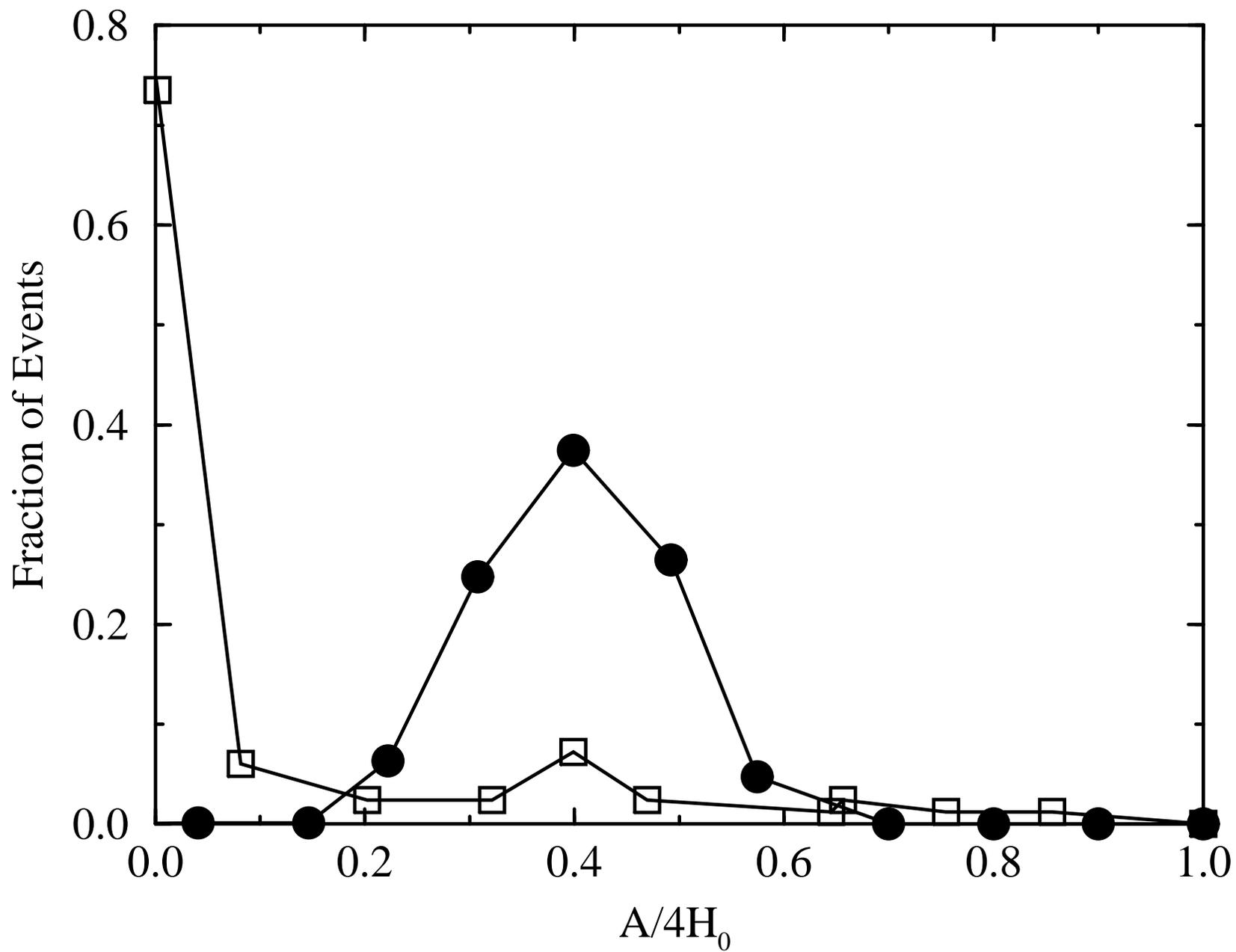
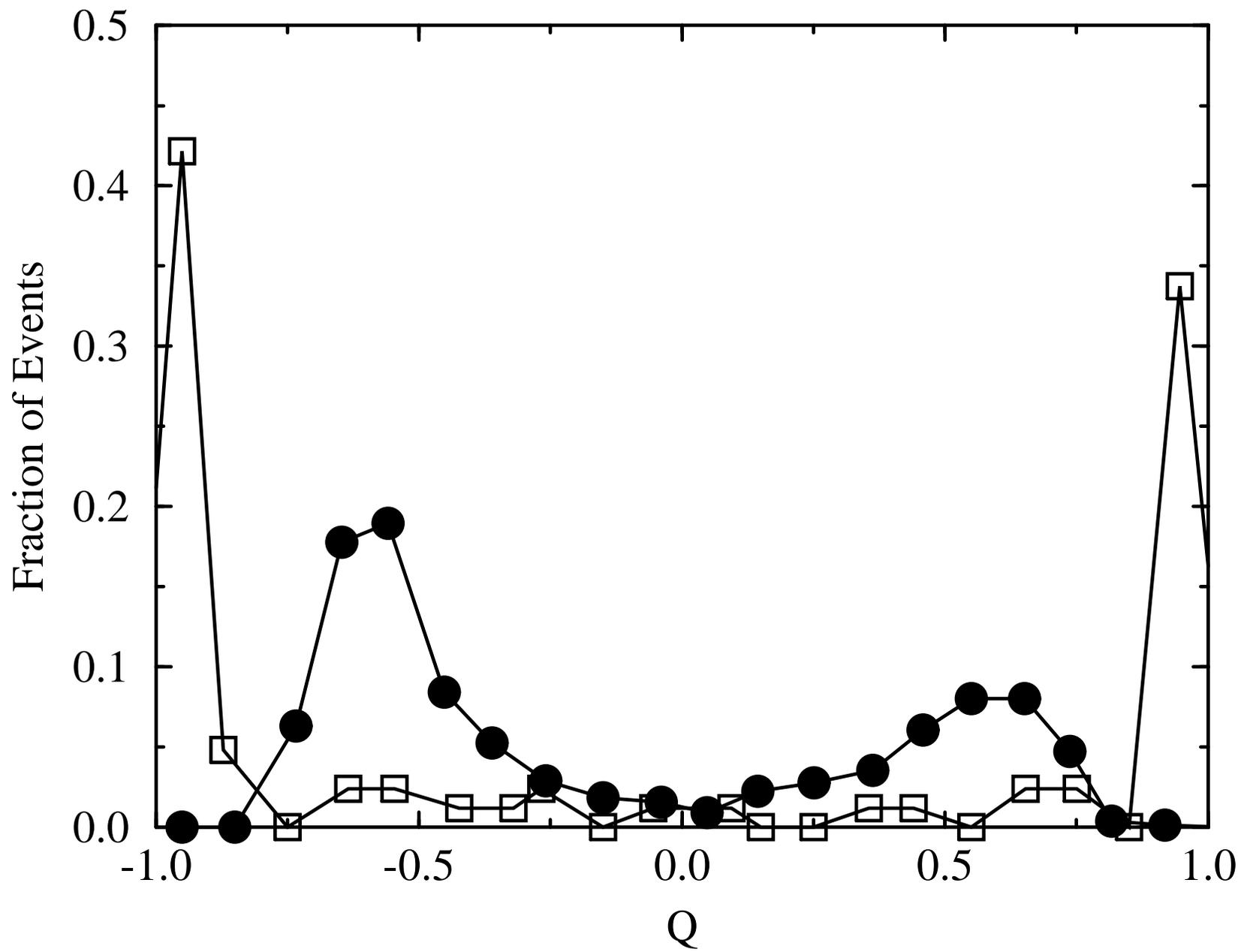
\end{document}